\documentclass[reprint,amsmath,amssymb,aps,pra]{revtex4-1}
\usepackage{graphicx}
\usepackage{dcolumn}
\usepackage{mathrsfs,amsmath}
\usepackage{bm}
\usepackage{xcolor}
\newcommand{\beq}{\begin{equation}}
\newcommand{\eneq}{\end{equation}}

\begin{document}

\title{Phase diagram of muonium hydride: when dimensionality matters}

\author{Jieru Hu$^\star$ and Massimo Boninsegni }
\affiliation{
Department of Physics, University of Alberta, Edmonton, T6G 2H5, Alberta, Canada}
\email{Present address: State Key Laboratory of Precision Spectroscopy, East China Normal University, 500 Dongchuan Road, 200241, Shanghai, China}
\date{\today}
\begin{abstract}
We carry out a theoretical investigation of the low-temperature phase diagram of muonium hydride in two dimensions, using numerical simulations. It is shown that the phase diagram of this substance 
is qualitatively different in two and three dimensions. Specifically, while in three dimensions it has been shown to be essentially identical to that of parahydrogen, i.e., only displaying a single (crystalline) phase, in two dimensions it is very similar to that of $^4$He, with an equilibrium liquid phase that turns superfluid  at a temperature as high as $\sim$ 2.2 K, and that crystallizes under applied pressure. To our knowledge, this is the first well-described case of a condensed matter system whose phase diagram is drastically altered by dimensional reduction.
\end{abstract}
\maketitle

\section{Introduction}
\label{intro}
One of the most important tenets of modern condensed matter theory, is that thermodynamic properties of macroscopic assemblies of particles are strongly affected by dimensionality. Far-reaching fundamental statements can be made about the thermodynamics of condensed matter systems in reduced dimensions, such as that no true long-range order can exist at any finite temperature (e.g., no Bose-Einstein condensation can take place). Moreover, phase changes like melting, or the superfluid transition, are amenable to very different theoretical descriptions, depending on whether they occur in three or two physical dimensions; in one dimension (1D), the very notion of phase transition becomes blurred. 
Furthermore, there exist remarkable physical phenomena that only occur if particle motion is restricted to fewer than three dimensions (3D). One may think of the Quantum Hall Effect, which owes its very existence to the confinement of electrons  to two dimensions (2D).  \\ \indent The observation of possible, novel behavior of condensed matter systems in which particle motion is confined to 2D or 1D has motivated much experimental research over the past few decades. This includes, for example, the study of atomically thin films of helium adsorbed on different substrates, or in the essentially one-dimensional confines of carbon nanotubes.
Lowering the dimensionality (e.g., from three to two), with the ensuing reduction in the magnitude of the potential energy of interaction among particles (due to the smaller coordination number), was also speculated to bring about important qualitative changes in the phase diagram of a substance, with respect to what is observed in 3D; this was especially thought to be the case for highly quantal fluids, such as helium or parahydrogen. An important possible consequence, in the case of parahydrogen, may be the stabilization of a liquid phase, turning superfluid at low temperature \cite{Ginzburg1972}, whose occurrence is preempted by crystallization in 3D (no supersolid or metastable superfluid phases exist) \cite{Boninsegni2018}. 
\\ \indent
If the interactions among particles are anisotropic (e.g., dipolar), then the phase diagrams in 2D and 3D can differ significantly \cite{Kora2019}; however, reliable theoretical studies have shown that, aside from well-understood, mainly formal (in this context) modifications to the theory brought about by dimensional reduction, the phase diagram in 2D of systems with central, pair-wise interactions remains qualitatively the same as in 3D.
For Lennard-Jones type systems such as helium, the basic features of the phase diagram are determined by the value of a dimensionless parameter of the form $\Lambda\equiv \hbar^2/(m\epsilon\sigma^2)$, where $m$ is the mass of the particles, while $\epsilon$ and $\sigma$ are, respectively, the depth of the attractive well and the diameter of the repulsive core at short distances of the pair potential \cite{Kora2020}.  
\\ \indent
To mention a few examples among systems for which quantum-mechanical effects are most significant, the phase diagram of the electron gas is qualitatively the same in 2D and 3D \cite{Drummond2009,Holzman2020}. The same is true of $^4$He, which in 2D has an equilibrium superfluid phase at low temperature, which crystallizes under pressure \cite{Gordillo1998}, and for parahydrogen, only displaying a crystalline phase \cite{Boninsegni2004}, with not even metastable superfluid phases, just as in 3D; in both cases, no qualitative change takes place, with respect to the three-dimensional case. The only ``deviation'' from this rule is represented by the lighter isotope of helium, namely $^3$He; in 3D, the equilibrium phase at low temperature is a self-bound (superfluid) liquid, whereas in 2D the system only exists in the gas phase \cite{Miller1978}. While this has significant implications, e.g., on the phase diagram of isotopic helium mixtures in 2D \cite{Hallock1998}, it can be regarded as a mainly quantitative change, given the structural similarity between the liquid and gas phases, which enjoy the same symmetry. 
\\ \indent
In this paper, we report the results of a theoretical investigation of a condensed matter system for which the dimensional reduction from 3D to 2D brings about a {\em major, qualitative} change in the phase diagram; to our knowledge, this is the first clear-cut example of such an occurrence. The substance is hypothetical, although in principle neither unphysical nor implausible, and it is named {\em muonium hydride}. A muonium hydride (HMu) molecule is a bound state of muonium and hydrogen atoms (a muonium atom is one in which a proton is replaced by an anti-muon). Its mass is only slightly more than half the mass of a parahydrogen molecule, and its electronic properties, in its ground state, are very nearly the same of parahydrogen \cite{Suffczynski2002,Zhou2005}. It has spin zero, and therefore a system of identical such particles obeys Bose statistics. \\ \indent 
The phase diagram of this system in 3D was studied in Ref. \onlinecite{Kora2021}, and was found to be identical to that of parahydrogen, albeit the ground state equilibrium crystalline phase has a remarkably low density, lower than that of $^4$He. Just like for parahydrogen, no evidence of any metastable (super)fluid phase can be found.
As shown below, in 2D the situation is entirely different. Our study, based on the same methodology used in Ref. \onlinecite{Kora2021}, namely Path Integral Monte Carlo simulations, shows that HMu has the same low-temperature phase diagram of $^4$He, with an equilibrium liquid phase with a density very close to that of $^4$He, turning superfluid at a temperature $T\sim 2$ K. Upon being compressed, the system freezes into a triangular crystal, with a freezing density slightly lower than that of $^4$He. 
\\ \indent
It is known that quantum-mechanical exchanges of identical particles play a pivotal role in upending the classical picture, which is not otherwise significantly altered by zero-point motion alone \cite{Boninsegni2012}; in this particular system, the reduction of dimensionality has the effect of rendering exchanges energetically competitive with respect to the potential energy of interaction. This is consistent with the observation made in Ref. \onlinecite{Kora2021} of an enhanced superfluid response in HMu clusters (comprising a few tens of molecules), compared to their parahydrogen counterparts.
\\ \indent
Leaving aside the actual experimental feasibility of the observation of the physics studied here, which seems at least presently out of reach, condensed HMu nonetheless represents an interesting, possibly useful ``theoretical data point'', in assessing quantitatively the importance of dimensionality on the thermodynamics of a condensed matter system.

\section{Model and methodology}
\label{modham}
We consider an assembly of $N$ HMu molecules, regarded as point-like, identical particles of mass $m$ and spin $S=0$, whose motion is restricted to two physical dimensions.
The system is enclosed in a rectangular cell of area $A=L_x\times L_y$, with periodic boundary conditions in the three directions, yielding a
density $n=N/A$. 
The quantum-mechanical many-body Hamiltonian reads as follows:
\begin{eqnarray}\label{u}
\hat H = - \lambda \sum_{i}\nabla^2_{i}+\sum_{i<j}v(r_{ij})
\end{eqnarray}
where the first (second) sum runs over all particles (pairs of particles), $\lambda\equiv\hbar^2/2m=21.63$ K\AA$^{2}$, $r_{ij}\equiv |{\bf r}_i-{\bf r}_j|$ and $v(r)$ denotes the pairwise interaction between two HMu molecules. Just as in Ref. \cite{Kora2021}, we model the interaction between two HMu molecules by means of the well-known Silvera-Goldman potential \cite{Silvera1978}, which is the most commonly utilized in microscopic studies of the condensed phase of parahydrogen. 
This choice of pair potential is based on the same considerations offered in Ref. \onlinecite{Kora2021}, namely {\em a}) the use of a quantitatively more accurate potential is not likely to affect the conclusions of our study in a significant way, and {\em b}) the comparison of the results with those of other calculations (i.e., those of Ref. \onlinecite{Kora2021} in 3D) allows us to zero in on the effect of the dimensional reduction alone.
\\ \indent
The low-temperature phase diagram of the thermodynamic system described by Eq.  (\ref{u}) as a function of  density and temperature has been studied in this work by means of  numerical (path integral Monte Carlo) simulations, based on the continuous-space Worm Algorithm \cite{Boninsegni2006,Boninsegni2006b}.  Since this technique is by now fairly well-established, and extensively described in the literature, we shall not review it here; we used the canonical variant of the algorithm, in which the number of particles $N$ is fixed \cite{Mezzacapo2006,Mezzacapo2007}.\\ \indent
Details of the simulation are  standard; we made use of the fourth-order approximation for the short imaginary time ($\tau$)  propagator (see, for instance, Ref. \onlinecite{Boninsegni2005}), and all of the results presented here are extrapolated to the $\tau\to 0$ limit. We generally found numerical estimates for structural and energetic properties of interest here, obtained with a value of the time step $\tau\sim 3.0\times 10^{-3}$ K$^{-1}$ to be indistinguishable from the extrapolated ones, within the statistical uncertainties of the calculation. 
We carried out simulations of systems comprising a number $N$ of particles equal to 36 and 144. The cell geometry was taken to be square (rectangular, accommodating a perfect triangular crystal) for simulations of the system in the liquid (crystalline) phase.
\\ \indent
Physical quantities of interest for the bulk calculations include the energy per particle and pressure as a function of density and temperature, i.e., the thermodynamic equation of state in the low-temperature limit. It is found that, within the statistical errors of our calculation, the results for most quantities remain unchanged below $T=1$ K, i.e., results quoted for this temperature can be regarded as essentially ground state estimates.
We estimated the contribution to the energy and the pressure arising from pairs of particles at distances greater than the largest distance allowed by the size of the simulation cell by approximating the pair correlation function $g(r)$ with 1 for greater distances. On comparing the results obtained for the two systems of different sizes studied here, we determined this to be a quantitatively accurate procedure. We computed the pair correlation function and the related static structure factor, in order to assess the presence of crystalline order, which can also be  detected through visual inspection of the imaginary-time paths. 
\\ \indent
 We probed for possible superfluid order through the direct calculation of the superfluid fraction using 
 the well-established winding number estimator \cite{Pollock1987}. In order to assess the propensity of the system to develop a superfluid response, and its proximity to a superfluid transition, we also rely on a more indirect criterion, namely we monitor the frequency of cycles of permutations of identical particles involving a significant fraction of the particles in the system. 
While there is no quantitative connection between permutation cycles and the superfluid fraction \cite{Mezzacapo2008}, a global superfluid phase requires exchanges of macroscopic numbers of particles (see, for instance, Ref. \onlinecite{Feynman1953}).

\section{results}
\label{res}
\begin{figure}[h]
\center
\includegraphics*[width=1 \linewidth]{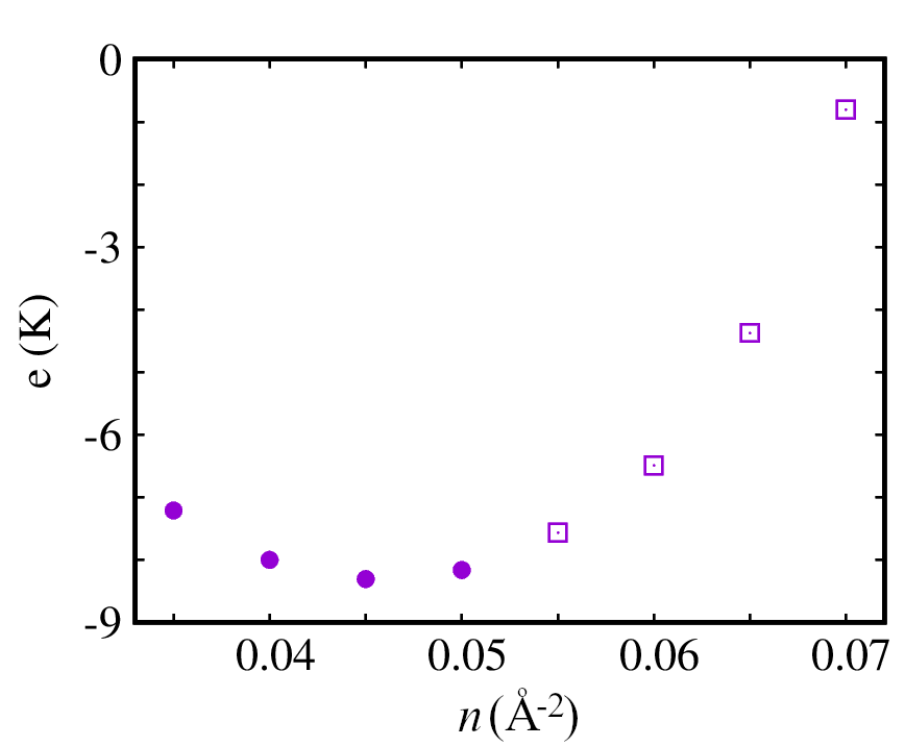}
\caption{ Energy per HMu molecules (in K) computed by simulation at a temperature $T=1$ K as a function of the density $n$ (in \AA$^{-2}$). Open squares refer to thermodynamic points at which the system is in the crystalline phase, filled circles to the superfluid. Statistical errors are smaller than symbol sizes.
}
\label{energy}
\end{figure}

We begin with a discussion of the low-temperature energetics of the system. Fig. \ref{energy}  shows the computed energy per molecule $e(n)$ at temperature $T=1$ K. As mentioned in \ref{modham}, the results remain unchanged, within the statistical errors of the calculation, or lowering the temperature below 1 K. Therefore, those shown in Fig. \ref{energy} can be considered as ground state estimates.
\\ \indent
The first observation is that the system is self-bound, i.e., there exists a nonzero equilibrium density $n_\circ\sim$ 0.045 \AA$^{-2}$; the binding energy $e_\circ \sim -8.5$ K \cite{note2}. The equilibrium density (magnitude of the binding energy) is approximately 2/3 (1/3) of that of parahydrogen in 2D \cite{Boninsegni2004}. Thus, on going from 3D to 2D  the binding energy is more significantly reduced  for HMu than parahydrogen. The value of $n_\circ$ is remarkably close to the equilibrium density in 2D of $^4$He \cite{Gordillo1998}. The question, of course, is whether the physics of the system at its equilibrium density is closer to that of $^4$He or parahydrogen. We address this issue by examining the structure and the superfluid response.
\begin{figure}[h]
\center
\includegraphics*[width=1 \linewidth]{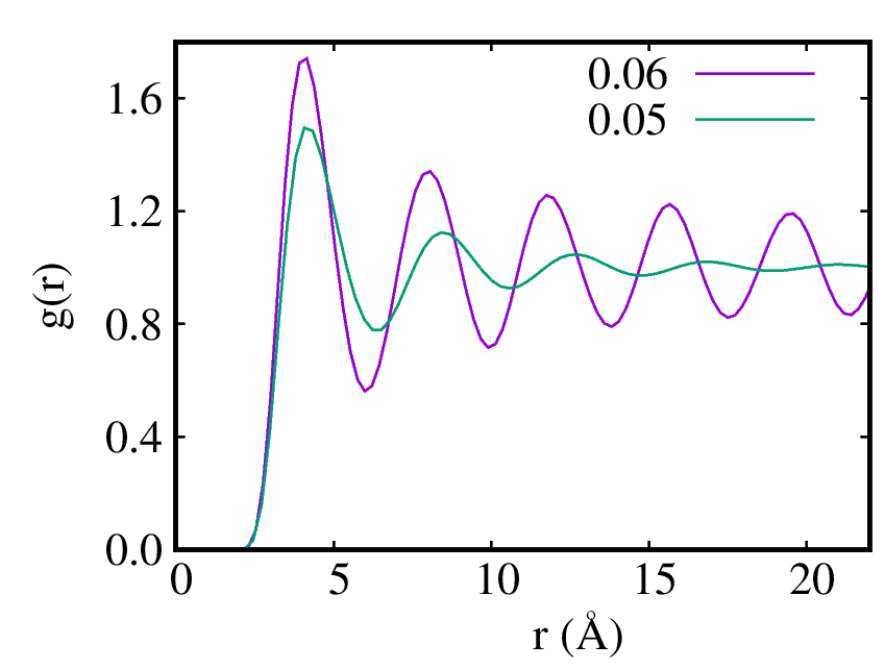}
\caption{ Pair correlation functions computed at $T=1$ K for the two different densities 0.05 and 0.06 \AA$^{-2}$.
}
\label{structure}
\end{figure}
\\ \indent
Fig. \ref{structure} shows the pair correlation function $g(r)$ computed by simulation at temperature $T=1$ K for the two different densities $n=0.05, 0.06$ \AA$^{-2}$. The qualitative difference between the two cases is clear, and it is important to remember that no {\em a priori} physical assumption is built into a finite temperature calculation, i.e., crystalline order, signaled by the regular oscillations in the $g(r)$ extending to long distances for the higher density, occurs spontaneously, not because of possible simulation bias which affects, for example, ground state Monte Carlo simulations \cite{Boninsegni2012b,Boninsegni2001}.
\\ \indent
At the lower value of the density, there is no crystalline order; on the other hand, long particle exchanges set in, involving a large fraction of all the particles in the system. As a result, off-diagonal (quasi) long-range order and a concomitant finite superfluid response appear. All of this is instead absent at the higher density, at which essentially no exchanges take place (no supersolid response is observed).
\begin{figure}[h]
\center
\includegraphics*[width=1 \linewidth]{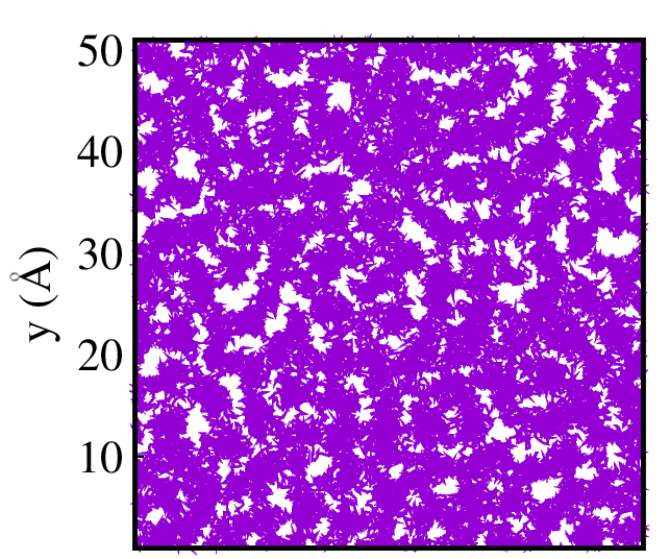}
\includegraphics*[width=1 \linewidth]{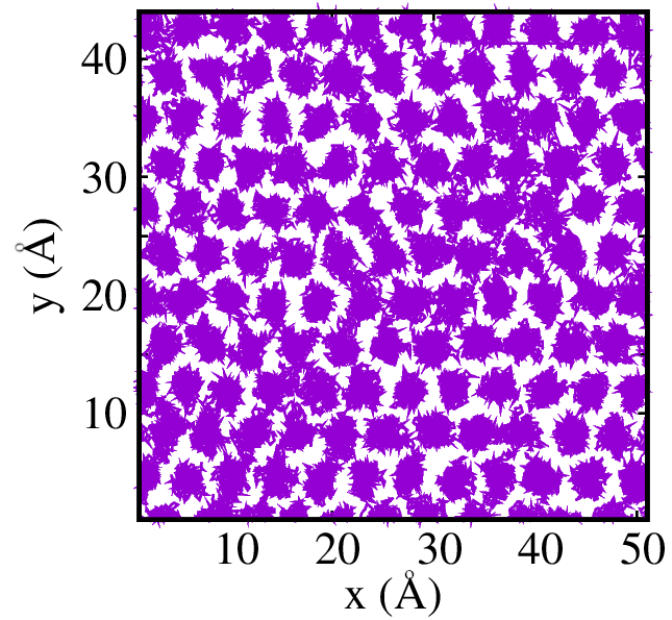}
\caption{Snapshots of many-particle worldlines for two representative configurations, top (bottom) illustrating the physics of the superfluid (crystalline) phase of density 0.05 (0.06) \AA$^{-2}$. 
}
\label{snap}
\end{figure}
\\ \indent
As mentioned above, additional, perhaps more intuitive insight can be provided by a direct, visual inspection of many-particle configurations generated by the sampling. An illustration is provided in Fig. \ref{snap}, which shows two such representative ``snapshots'', the top (bottom) for the system at density 0.05 (0.06) \AA$^{-2}$. In the top snapshot, which refers to the lower density, particles are delocalized and their worldlines are highly entangled, i.e., exchanges of identical particles occur frequently \cite{Feynman1953}. 
\\ \indent
This can be contrasted with what happens in the crystalline phase (bottom panel of Fig. \ref{snap}), in which particles are localized at lattice sites; exchanges occur infrequently and remain local in character, hence no global superfluid response arises. Indeed, no ``supersolid'' behavior is observed in HMu, just as in parahydrogen, consistent with the current understanding of supersolidity which requires that the interaction among particles feature a ``soft'' repulsive core at short distances \cite{Boninsegni2012c}.
\begin{figure}[h]
\bigskip
\center
\includegraphics*[width=1 \linewidth]{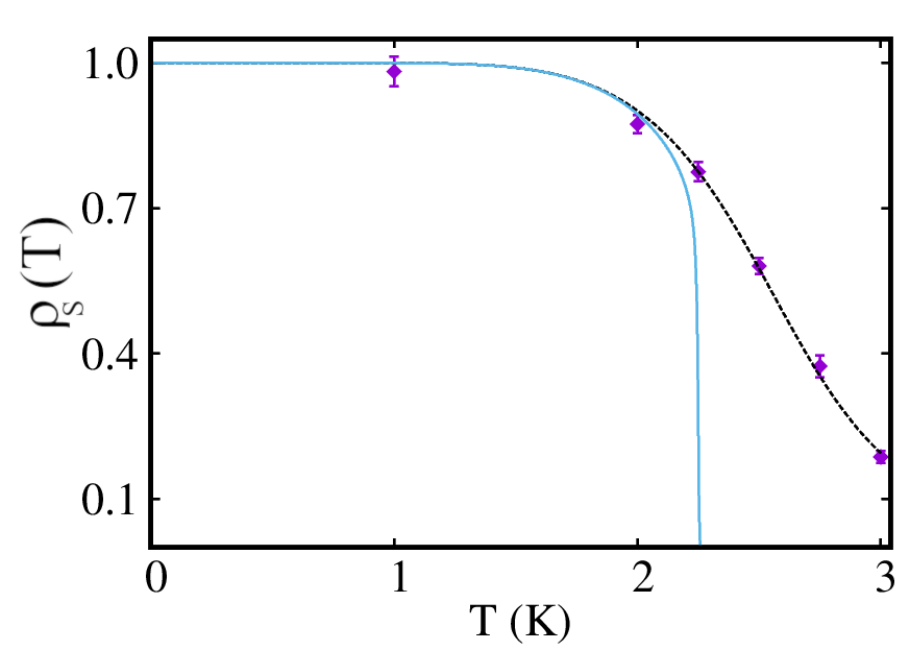}
\caption{ Superfluid fraction $\rho_S(T)$ as a function of temperature, for a system comprising $N=144$ HMu molecules. Diamonds are values computed by simulation, the dotted line is a fit to the data based on the BKT recursive equations (see text), while the solid line represents the extrapolation to the thermodynamic limit. When not shown, statistical errors are smaller than the symbol size.
}
\label{superfluid}
\end{figure}
\\ \indent
The superfluid transition that takes place in the fluid phase conforms to the well-known Berezinskii-Kosterlitz-Thouless paradigm. Fig. \ref{superfluid} displays the estimates for the superfluid fraction $\rho_S(T)$
 as a function of temperature, calculated by simulation for a system comprising $N=144$ HMu molecules.
 K.  The fit to the data is obtained following the procedure outlined in Ref. \onlinecite{Ceperley1989}. Based on the fit to the data for the finite system, one can infer the behavior in the thermodynamic limit, shown by the solid curve in Fig. \ref{superfluid}, and with the aid of the well-known universal jump condition \cite{Nelson1977}, we estimate the superfluid transition temperature $T_c$ to be approximately 2.2
 K. This can be compared to the value of $T_c\approx 0.65$ K for $^4$He in 2D. The enhancement of the critical temperature of HMu, with respect to that of $^4$He, arises from the combination of lower mass and stronger interaction, in turn resulting in a relatively high equilibrium density.

\section{Conclusions}
We have investigated the low-temperature phase diagram of a bulk muonium hydride in 2D, by means of unbiased, reliable quantum simulations. Our model assumes a pairwise, central interaction among HMu molecules identical to that of H$_2$ molecules. Quantitative arguments have been offered in Ref. \onlinecite{Kora2021} in support of this assumption; moreover the use of the same potential utilized in Ref. \onlinecite{Kora2021} allows one to isolate the effect of the dimensional reduction (from 3D to 2D), which is the main subject of our study.
\\ \indent
The significant (in our view) outcome of this study is the radically different physics that the system displays in 2D and 3D. To our knowledge, this is the first clear-cut example of dimensional reduction altering the phase bulk phase diagram of a substance as observed in this work. While in 3D HMu is a crystal, and its phase diagram is virtually identical to that of parahydrogen (albeit with a remarkably low equilibrium density), in 2D its phase diagram mimics that of $^4$He, with a ground state superfluid equilibrium phase which crystallizes only on compression, and with a remarkably high critical temperature, above 2 K. In other words, the physical scenario once deemed possible, but which is {\em not} realized for parahydrogen, is predicted to take place for muonium hydride, at least as described by the model considered here.
\\ \indent
The experimental feasibility of the system considered here appears severely hampered by the very short lifetime of the $\mu^+$ (of the order of a $\mu$s), replacing one of the two protons of a parahydrogen molecule, resulting in one of muonium hydride. To be sure, the production of a single, or few HMu molecules has already been demonstrated; indeed, ``muonium chemistry'' has been a legitimate, active area of research for quite some time \cite{muonium}. Clearly, however, assembling a sufficiently large number of such objects in some restricted region of space, in order to carry out meaningful observations of condensed matter effects (a challenge in many respects reminiscent of those faced by cold atom experimenters), seems a truly daunting task, at the present time. 
\\ \indent
Still, even if it were to be regarded exclusively as a ``toy'' system, HMu can still prove a useful theoretical construct, both as a limiting case as well as offering nontrivial insight into the interplay between superfluid and crystalline orders. Of particular interest, for example, is the study (e.g., by means of computer simulation) of layering, and the emergence of solid order, as successive superfluid layers of HMu are adsorbed on a substrate.
\label{concl}

\section*{Acknowledgements} 
This work was supported by the Natural Sciences and Engineering Research Council of Canada, and by the China Scholarship Council.

   \bibliography{refs}

\begin{thebibliography}{30}
\bibitem [{not()}]{note2}%
  \BibitemOpen
  \href@noop {} {}\bibinfo {note} {An accurate, quantitative characterization
  of the phase diagram, including the calculation of the binding energy and the
  equilibrium density, as well as a precise determination of phase boundaries,
  was not pursued in this work. The reasons are that it is not relevant to the
  main physical conclusions made here, and it is not of special significance
  considering the uncertainty in the interaction potential.}\BibitemShut
  {Stop}%
\end{thebibliography}
\end{document}